%% file: main.tex
\documentclass[runningheads]{llncs}
\usepackage[T1]{fontenc}
\usepackage{graphicx}
\usepackage{amssymb}

\input{preamble}

\begin{document}
\title{Statically Inferring Usage Bounds for Infrastructure as Code
}
%
%
\author{Feitong Qiao\inst{1} \and
    Aryana Mohammadi\inst{2} \and
    Jürgen Cito\inst{3}\orcidID{0000-0001-8619-1271} \and
    Mark Santolucito\inst{2}\orcidID{0000-0001-8646-4364}}
\authorrunning{F. Qiao et al.}
%
\institute{Columbia University, New York NY, USA
    \and
    Barnard College, Columbia University, New York NY, USA 
    \and
    TU Wien, Vienna, Austria
}
\maketitle              
\begin{abstract}
    Infrastructure as Code (IaC) has enabled cloud customers to have more agility in creating and modifying complex deployments of cloud-provisioned resources.
    By writing a configuration in IaC languages such as CloudFormation, users can declaratively specify their infrastructure and CloudFormation will handle the creation of the resources.
    However, understanding the complexity of IaC deployments has emerged as an unsolved issue.
    In particular, estimating the cost of an IaC deployment requires estimating the future usage and pricing models of every cloud resource in the deployment.
    Gaining transparency into predicted usage/costs is a leading challenge in cloud management.
    Existing work either relies on historical usage metrics to predict cost or on coarse-grain static analysis that ignores interactions between resources.
    Our key insight is that the topology of an IaC deployment imposes constraints on the usage of each resource, and we can formalize and automate the reasoning on constraints by using an SMT solver.
    This allows customers to have formal guarantees on the bounds of their cloud usage.
    We propose a tool for fine-grained static usage analysis that works by modeling the inter-resource interactions in an IaC deployment as a set of SMT constraints, and evaluate our tool on a benchmark of over 1000 real world IaC configurations.
\end{abstract}
\input{secs/intro}
\input{secs/example}
\input{secs/system_overview}
\input{secs/experiment}

\input{secs/related}
\input{secs/discussion}

\subsubsection{Acknowledgements} 

This work has been supported by the National
Science Foundation under Grant No. CCF-2105208 and by the European Union's Horizon research and innovation programme under grant agreement No. 101086248 (Cloudstars).
\newpage
%
%
%

%

\bibliographystyle{splncs04}
\bibliography{references}

%
%
%
%
%
\end{document}

%% file: preamble.tex
\usepackage{listings}
\usepackage{xspace}
\usepackage{amsmath}
\usepackage{subcaption}

\usepackage{todonotes} 
\setuptodonotes{inline}

\usepackage{cite}

%% file: secs/intro.tex
\section{Introduction}

One of the most pressing issues with IaC deployments is in the difficulty of estimating pricing~\cite{anodot2022}.
Despite cloud providers' thorough documentation of pricing models and strong tool support, challenges remain in understanding the cost of large deployments.
In a recent industry survey, it was found that 49\% of IT executives find it difficult to get cloud costs under control, and 54\% of those believe a primary challenge is a lack of visibility into cloud usage~\cite{anodot2022}.
The issue of IaC analysis has also been recognized by the academic community to be of critical significance~\cite{cito_et_al:DagRep.13.2.163, cito:15, bohme:23}.

We identify two categories of existing tools for IaC cost estimation - those that rely on dynamic analysis and those that rely on static analysis.
Dynamic analysis tools fall short in capturing topological changes to infrastructure,
and existing static analysis tools require too much guesswork from the user.

AWS's Cost Explorer is one example of a dynamic analysis tool for IaC pricing.
This tool allows users to track the ongoing usage and costs incurred of all resources in a live CloudFormation deployment.
While dynamic analysis of IaC is helpful for existing infrastructure, such tools cannot be used for new deployments or topological modifications to existing infrastructure.
Even a small change to the topology of an infrastructure might redirect user requests through a different path of the IaC resource graph, rendering past resource specific usage patterns irrelevant as topological changes induce a change in the dataflow.

AWS's Pricing Calculator, on the other hand, is one example of a static analysis tool for IaC pricing.
This tool allows users to load a CloudFormation file, provide usage estimates, and see the anticipated cost of the overall infrastructure.
However, estimating usage is extremely difficult and a regular pain point for customers.
If the user is creating a new infrastructure, how can they estimate the usage? If there is a significant change to the topology of an existing infrastructure, how does the user know the extent to which past usage data can be extrapolated to the new infrastructure?

\input{figs/motiv}

At a high level, we propose a static analysis method for IaC configuration files that assists users in making correct usage estimates.
Our approach is, at its core, a modeling of the graph of the IaC and propagating local constraints across edges and nodes to be able to check global constraints.
The key contributions of this work are:

\begin{enumerate}
    \item A method for modeling the resource usage of cloud infrastructures as a set of composable SMT constraints, allowing for user queries about validity of usage estimates and usage bounds of the overall IaC stack;
    \item An implementation of this tool and an evaluation on a benchmark dataset of over 1000 real world IaC files.
\end{enumerate}

%% file: figs/motiv.tex
\begin{figure}[t!]
    \centering
    \begin{subfigure}{0.6\textwidth}
        \centering
        \includegraphics[width=.95\textwidth]{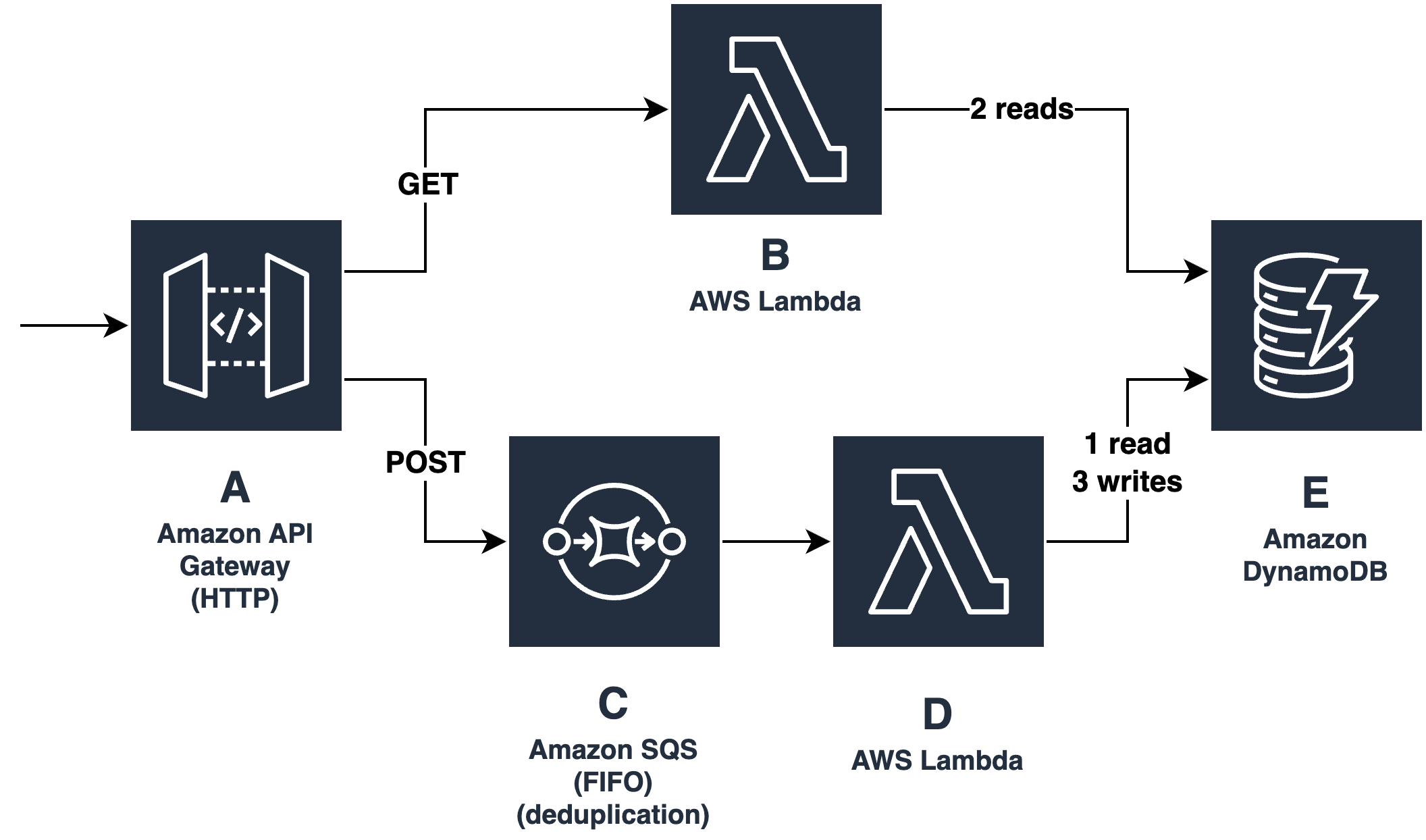}
        \caption{An example CloudFormation stack topology}
        \label{fig:example}
    \end{subfigure}%
    \begin{subfigure}{0.4\textwidth}
        \centering
        {\footnotesize
            \begin{align*}
                \\ \\
                 & A.POST + A.GET = 1000000      \\
                 & 0 \leq w \leq 3 * A.POST      \\
                 & 0 < r \leq A.POST + 2 * A.GET \\
                 \\ 
            \end{align*}}%
        \caption{Constraints on usage}
        \label{fig:equations}
    \end{subfigure}
    \caption{Motivating example}
\end{figure}

%% file: secs/example.tex
\section{Motivating Example}
\label{sec:example}

As an illustrative example, consider the infrastructure depicted in Figure \ref{fig:example}.
Assuming that this infrastructure is new, there will be no available historical usage data.
When a user wants to estimate the cost associated with this infrastructure, they must provide estimates for the utilization of each resource.
Generating these estimates requires the user to manually infer the relationship between resources, such as the correlation between (A) and (E), based on the infrastructure's topology.

For instance, if we envision 1 million requests on resource (A), then the number of read requests ($r$) and write requests ($w$) on resource (E) must adhere to the system of inequalities listed in Fig.~\ref{fig:equations}.
Correctly inferring these bounds requires an understanding of how every resource propagates requests through the graph, while accounting for alterations in the quantity and types of requests.
Inferring this set of constraints is a non-trivial task and requires significant experience and familiarity with all the configuration options of every resource.
This process only becomes more difficult with larger and more complex IaC configurations.

%% file: secs/system_overview.tex
\section{System Overview}

\begin{figure}[t]
    \centering
    \includegraphics[width=\textwidth]{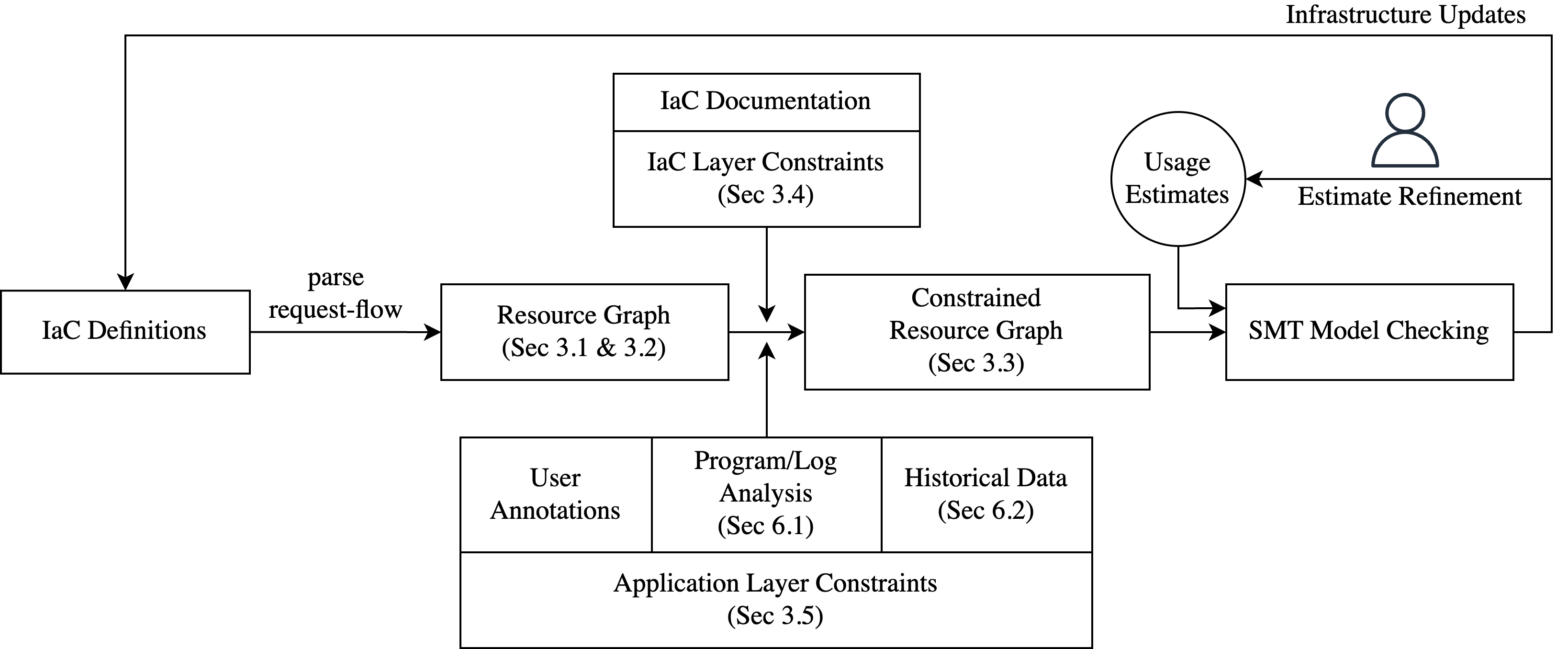}
    \caption{System diagram}
    \label{fig:systemdiagram}
\end{figure}

As shown in Fig.~\ref{fig:systemdiagram}, our system's input is an IaC configuration provided by the user. The tool uses this to build a resource graph, instantiates metric variables that represent the resource usage measurements, and generates constraints that relate the metrics variables and model the dataflow behaviors within the infrastructure. It is currently implemented for the AWS CloudFormation IaC platform targeting the AWS cloud platform, but the process can be similarly applied to other IaC languages (e.g. Terraform) and cloud platforms (e.g. GCP).

\subsection{Building the resource graph}

The major cloud platforms provide a large variety of resource types, and each resource type may have its own sets of usage measurement metrics. We will use $RT$ to denote the set of resource types.

For each resource type, there may be public and private metrics that together model the resource usage measurements.
For example, when reasoning about the usage of an AWS SQS within an infrastructure, each request/message within the SQS triggers a downstream request, therefore it is natural for programmers to think in terms of the number of requests/messages within the queue.
But on the other hand, the pricing system meters SQS usage using the number of API calls to the SQS.
To take both of these metrics into account when modeling the system, the tool instantiates a metric variable for each of them, with the first being a public metric variable and the second being private.
This difference is not just nominal and will be discussed in Sec. \ref{sec:constraint_generation}.

We denote $M$ to be the set of all metric names (e.g. \textit{"monthly\_requests"}), and use $metrics_{pub} : RT \rightarrow 2^{M}$ and $metrics_{priv}: RT \rightarrow 2^{M} $ to denote the assignments from a resource type to its set of public/private metrics respectively. We also define $metrics(t) = metrics_{pub}(t) \cup metrics_{priv}(t)$ to be the set of both public and private metrics for any given resource type. Metrics can be either integers or real numbers.

From the topology defined in the IaC definitions, we can build a resource graph, where the nodes $N$ are the resources and the directed edges $E \subseteq N \times N$ are direct connections/triggers. We note that the resource graph is a dataflow graph instead of a dependency graph. Unlike a IaC dependency graph in which a directed edge corresponds to a partial order of deployment~\cite{lepiller2021analyzing}, a directed edge in the resource graph models how a resource induces requests to specific other resources. For later use, we also define the function $rtype : N \rightarrow RT$ that retrieves each node's resource type.

\subsection{Node variables and edge variables}

After building the resource graph from the dataflow analysis, we annotate the resource graph with metric variables.
To model the quantitative details of the dataflow behaviors within the infrastructure, we instantiate \textit{node variables} and \textit{edge variables}.

A \textit{node variable} represents a single metric of a resource. The set of all node variables $NV$ has a one-to-one correspondence to $\{ (n, m) | n \in N, m \in metrics(rtype(n)) \}$. In other words, for each node, we instantiate a node variable for each metric associated with its resource type. We also use $NV_{pub}, NV_{priv} \subseteq NV$ to denote the subset of public node variables (node variables for public metrics) and the subset of private node variables (node variables for private metrics).

An \textit{edge variable} represents the value of a public metric that a node receivs from its predecessor. The set of all edge variables $EV$ has a one-to-one correspond correspondence to $\{ (s, t, m) | (s, t) \in E, m \in metrics_{pub}(rtype(t)) \}$. In other words, for each edge, we instantiate an edge variable for each public metric associated with the destination node.

To facilitate the constraint generation step next, for each node $n_0$, we let

\begin{itemize}
    \item $NV(n) = \{ v | v = (n, m) \in NV, n = n_0 \}$ denote the set of node variables for $n_0$;
    \item $NV_{pub}(n_0) = \{ v | v = (n, m) \in NV_{pub}, n = n_0 \}$ denote the set of public node variables for $n_0$;
    \item $NV_{priv}(n_0) = \{ v | v = (n, m) \in NV_{priv}, n = n_0 \}$ denote the set of private node variables for $n_0$;
    \item $EV_{in}(n_0) = \{ v | v = (s, t, m) \in EV, t = n_0 \}$ denote the set of \textit{incoming edge variables}, i.e. the edge variables for which $n_0$ is the source node;
    \item $EV_{out}(n_0) = \{ v | v = (s, t, m) \in EV, s = n_0 \}$ denote the set of \textit{outgoing edge variables}, i.e. the edge variables for which $n_0$ is the destination node.
\end{itemize}

\subsection{Constraint generation}
\label{sec:constraint_generation}

A constraint $\phi$ is an SMT formula such that $FV(\phi) \subseteq NV \cup EV$, where $FV(\phi)$ denotes the set of variables that occur free in $\phi$.
For each variable $v \in NV \cup EV$, there may be zero or more \textit{basic} constraints that describe basic properties of the metric.
At the moment, all supported resources in our tool have just the basic constraints that each metric should be greater or equal to 0.
More may need to be added as the tool supports more resources and metrics.

Next, for each node $n \in N$, the tool generates zero or more constraints. Each of these constraints $\phi \in \mathcal{C}$ is categorized as one of the following:
\begin{enumerate}
    \item \textbf{incoming}: for a public node variable $nv \in NV_{pub}(n)$, the constraint $\phi$ relates $nv$ to the incoming edge variables, i.e. $FV(\phi) \subseteq \{nv\} \cup EV_{in}(n)$
    \item \textbf{intrinsic}: the constraint $\phi$ relates node $n$'s node variables to each other, i.e. $FV(\phi) \subseteq NV(n)$
    \item \textbf{outgoing}: for an outgoing edge variable $ev \in EV_{out}(n)$, the constraint $\phi$ relates $ev$ to its public node variables, i.e. $FV(\phi) \subseteq NV(n) \cup \{ev\}$.
\end{enumerate}

Intrinsic constraints capture intra-resource behavior.
For example, in the motivating example Fig. \ref{fig:example}, all incoming requests to (A) should be greater than the sum of the POST and GET requests to (A).
Written as an SMT constraint, $\textit{A.monthly\_requests} = \textit{A.monthly\_GETs} + \textit{A.monthly\_POSTs}$.

Incoming and outgoing constraints capture inter-resource behavior.
For the motivating example from Sec.~\ref{sec:example}, the number of POST requests to (A) should equal the number of monthly requests to (C).
Written as an SMT constraint, the outgoing constraint from A to C is $\textit{A.monthly\_GETs} = \textit{A\_C.monthly\_requests}$.
These are more difficult to derive as they can arise from both the IaC level configuration as well as the application layer (discussed in Sec.~\ref{sec:iacLayer} and Sec.~\ref{sec:appLayer}).

These three types of constraints allow the tool to model the system globally with only node-local knowledge.
When generating constraints for a node, there is no need for graph traversal of any kind.
The global set of constraints can be collected by simply iterating through the nodes.

\subsection{IaC Layer Constraints}
\label{sec:iacLayer}
For many modern cloud serverless resources (e.g. AWS SQS and APIGateway), our tool is able to generate all the necessary constraints using the information within the CloudFormation templates. Take our motivating example from Sec.~\ref{sec:example}. The node (C) is an SQS service with FIFO and deduplication enabled. By consulting the AWS documentation, we know that with these configurations, the requests that come out of (C) (and in this graph are sent to (D)) will be less than or equal to the requests sent to SQS. This induces an outgoing constraint $\textit{C.monthly\_requests} >= \textit{C\_D.monthly\_requests}$. This constraint is an \textit{IaC layer} constraint and can be derived purely from the IaC code.
Every cloud resource is given a SMT constraint template, as manually derived from the documentation, that is used to generate these constraints.

\subsection{Application Layer Constraints}
\label{sec:appLayer}
For cloud resources with programmable behaviors, additional knowledge about the application program allows us to generate constraints that are not discoverable with just the IaC definitions.
We call these \textit{application layer} constraints.
For example, in the motivating example, we have a constraint on the edge from (D) to (E) that the Lambda will induce 1 read and 3 writes to the DynamoDB (E) for every request to the Lambda.
However, the information needed to generate such a constraint is beyond the scope of the IaC configuration, and thus our tool is currently not able to infer application layer constraints.
Our tool instead relies on user-provided custom SMT constraints for application layer constraints.
In Sec.~\ref{sec:discussion_appLayer}, we discuss some approaches one may take to automate the inference of application layer constraints.

%% file: secs/experiment.tex
\section{Evaluation}

\textbf{Methodology} We draw our benchmark set from the PIPr dataset of public IaC programs~\cite{Sokolowski_Spielmann_Salvaneschi_2023}. PIPr contains 7104 public repositories of Programming Languages Infrastructure as Code (PL-IaC) projects written with Pulumi, AWS CDK or Terraform. With automated scripts (provided in our GitHub repository\footnote{\url{https://github.com/Barnard-PL-Labs/IaCAnalysis}}), we identify the AWS CDK projects implemented in JavaScript or TypeScript, and run `\textit{npm install; cdk synth}' in each project to synthesize a CloudFormation template for each. The result is our benchmark set containing 1062 CloudFormation templates.

\subsection{Quantitative Analysis}

The tool currently supports 12 AWS CloudFormation resource types.
The tool also flags 22 resource types as non-dataflow related and omits these in the construction of the resource graph.
A resource type is non-dataflow related if it is not an infrastructure resource (for instance, IAM policies and Lambda permissions).
Each supported resource type typically has 2 metrics.

As a tool designed to be integrated into an existing user workflow, our tool completes in a short amount of time, spending only $0.5 \pm 0.15$ seconds for all samples on a 32GB M1 Max MacBook Pro.
As a proxy measure of complexity, we looked at the 457 templates that have at least 3 supported resource nodes, and found that the average graph degree (i.e. number of edges directed into a node) is 0.9 with a standard deviation of 0.6.
This means that IaC resource graphs are usually very sparse and thus gives us confidence that our tool will scale well, even with larger IaC configurations.
Combined with the finding that the number of resources is usually small (benchmark set averaging $21.9$ with standard deviation of $21.1$), we don't expect performance to be an issue even on IaC configurations beyond our benchmark set.

\begin{figure}
    \includegraphics[width=1\textwidth]{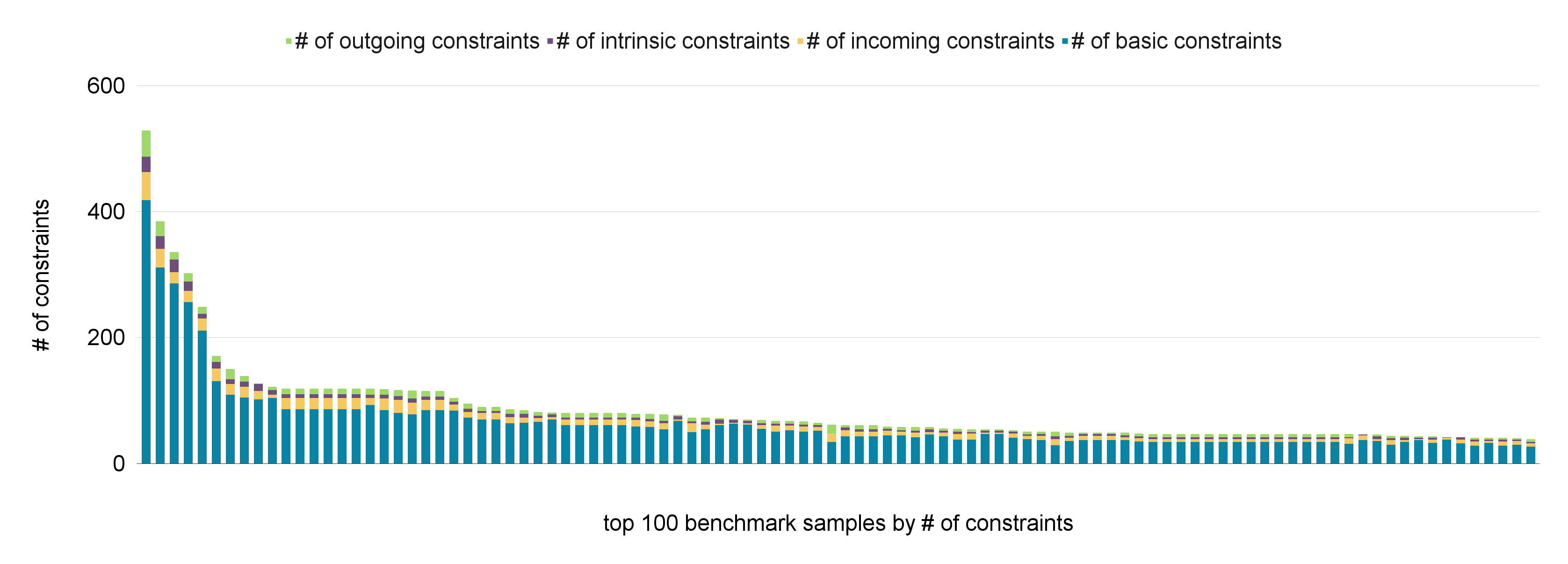}
    \caption{Number of constraints for the benchmark samples (top 100).}
    \label{fig:constraints_chart}
\end{figure}

We ranked the samples by the number of constraints generated by the tool, and show the top 100 samples in Fig. \ref{fig:constraints_chart}. 
Each bar is also broken down into the four constraint types: basic, incoming, intrinsic and outgoing.
This chart shows that a small number of IaC configurations have significant complexity (benchmark \#1 in Fig.~\ref{fig:constraints_chart} has 529 constraints), and there is a long tail of IaC configurations that still have enough complexity such that there is value in automating this reasoning task (benchmark \#100 in Fig.~\ref{fig:constraints_chart} has 39 constraints).

\subsection{Case Study}

To concretely demonstrate the workflow of our tool `\texttt{iac-analysis}', we walk through an illustrative example of a developer tasked with implementing and deploying a simple project.
The project is sampled from the PIPr dataset, with ID 501027421, and is a public GitHub repository named "HektorCyC/url-shortener-app".
Written in TypeScript and AWS CDK, the project is a URL shortener application, in the form of a chain from APIGateway to Lambda to DynamoDB.
When run through our tool, this IaC configuration has 28 constraints (ranking \#138 in the benchmarks, sorted by number of constraints).

After coding up the project, the developer synthesizes a CloudFormation template named \texttt{cfn.yaml} and runs \texttt{iac-analysis estimates-template cfn.yaml}. This command provides them with a template for usage estimates. After being filled in, the template looks as following:

\begin{center}
    \begin{tabular}{c}
        \begin{lstlisting}
apigateway:
  monthly_GETs: 100000000
  monthly_PATCHs: 100000000
  monthly_DELETEs: 100000000
dynamodb_table:
  monthly_dynamodb_r: 3000000
        \end{lstlisting}
    \end{tabular}
\end{center}

The Lambda application receives GET, PATCH and DELETE requests from the APIGateway. For a GET request, it reads the DynamoDB once; for a PATCH or DELETE request, it reads once and writes once to the DynamoDB. The developer writes these application layer constraints as custom SMT-LIB constraints in a file named \texttt{app\_constraints.smt2}, and runs \texttt{iac-analysis check cfn.yaml app\_constraints.smt2}. In the output, the tool reports that the user-provided estimates are invalid. The developer reviews the estimates again, discovering that they made a mistake and left out some zeros in the \texttt{monthly\_dynamodb\_r} usage estimate.

%% file: secs/related.tex
\section{Related Work}

Our work is related to cost analysis of cloud deployments and workload characteriztion.
Workload characterization has been an active research area with a focus on understanding resource utilization, performance, and cost implications~\cite{workloadsurvey2016}. Traditional approaches to workload characterization often involve profiling applications or services to understand their resource requirements and performance characteristics. These studies typically rely on statistical analysis of historical data \cite{azmandian2011workload, wolski2013using}.
Some existing works aim at dynamic cost estimation by considering real-time resource utilization metrics and billing information. These approaches often leverage machine learning techniques to predict costs based on historical usage patterns \cite{gao2020machine, saxena2023performance, dezhabad2019cloud}. However, they struggle to capture the intricate relationships and dependencies between resources defined in IaC deployments.
Recent research has explored topology-based cost models for understanding and optimizing cloud deployments~\cite{obetz2019static, lin2022fine, leitner2016modelling}. These approaches consider the arrangement and relationships between different components in an IaC deployment. However, they do not explicitly model the constraints imposed by the deployment topology and the component characteristics on resource usage.
Our proposed system instead builds a topology-based resource graph to enforce constraints inherent in the component (intrinsic contraints) and the relationship between components (incoming and outgoing constraints), providing end users with formal guarantees on usage bounds.

Unlike existing approaches that predominantly rely on accurate usage estimates to provide cost estimation in cloud deployments, our system focuses on evaluating the feasibility of such estimates by modeling the intricate relationships between infrastructure components in the form of a resource graph. By employing SMT constraints to encapsulate these inter-resource interactions, our approach allows for a more rigorous examination of the deployment's constraints, informing end users whether the estimated resource usage aligns with the inherent limitations and relationships defined within the infrastructure components. This novel perspective does not only enable existing approaches to predict costs more accurately but also to ascertain the validity and appropriateness of their initial usage estimates given the constraints imposed by the deployment's topology.



%% file: secs/discussion.tex
\section{Discussion}

\subsection{Limitations}

A limitation of the evaluation is the lack of usage estimates in the benchmark.
The PIPr dataset provides sufficient IaC definition samples, but does not provide corresponding resource usage estimates.
Therefore, the tool is not benchmarked in its capability to check them against the generated constraints.
In fact, usage estimates are highly proprietary data, so they are generally not publicly available.
However, we don't expect validating the usage estimates against the generated constraints to be an expensive problem.

A few software engineering issues also still need to be resolved for the tool to be adopted in industrial workflows.
Often, especially in enterprise environments, IaC deployments are modularized.
Developing an application from multiple stacks is a common and even best practice in modern IT operations~\cite{koskelin2023}.
However, our tool requires a full view of the application's architecture to provide a comprehensive set of constraints.
One workaround is to create a merged template strictly for running cost estimates.
Another is to constrain individual templates and manually supply constraints that would connect the models together.
As a novel approach to cloud resource usage analysis, a user study would help to find the to most effective ways to integrate the tool into existing usage analysis workflows.



\subsection{Application Layer Constraints from Program Analysis}
\label{sec:discussion_appLayer}

For resources with programmable behaviors (e.g. AWS Lambda), the IaC definitions are usually paired with the application source code.
To extract the application layer constraints, an option is to run symbolic execution on the application source code.
This might be tractable if the application is a relatively small code snippet, but will be more difficult if the program logic is complex.
In the case that code analysis is not feasible, if we are updating an existing infrastructure and have log data (e.g. from the AWS Cost Explorer), we can infer some relation between the number of requests from one resource to the next.
In this case, we can check the logs and see the timestamped relations between incoming requests on the neighboring nodes.
The local relations on usage can be more reliably pulled from log data than global constraints, as the global relations depend on the topology of the infrastructure.

In the current version, the tool focuses on the Iac layer and allows user to submit the application layer constraints as custom SMT-LIB assertions.
This allows the tool to remain useful for infrastructures that utilize unsupported resources.
Take the motivating example. If we replace resource (B) with an EC2 instance, the tool cannot generate estimates for the entire architecture because that would require understanding the application runtime behavior.
However, the user may understand that behavior and can insert corresponding constraints.
Even in this case, the tool still greatly reduces the manual work and room for human error by taking care of the rest of the architecture.

\subsection{Reusing Historical Usage Data}

The static analysis and usage estimate procedure described above gives user the ability to put constraints on their infrastructure usage.
However, for updates to existing infrastructure, historical usage may be available and valuable.
We can incorporate this valuable data into our procedure by identifying subgraphs of the infrastructure that are minimally impacted by the IaC topological changes.

In particular, when processing an IaC update, any time we identify a node in our directed resource graph for which the historical data is not directly transferable (due to topological changes), we also mark all downstream nodes as having non-transferable historical data.
We would like to further explore opportunities to utilize this historical data (for example by migrating the data with respect to the updated constraints) in the new infrastructure.

\subsection{Conclusions}

We introduce a new method for modeling the resource usage of cloud infrastructures as a set of composable SMT constraints.
The generated set of constraints can be used to validate user-provided usage estimates and provide usage bounds of the overall IaC stack.
We provide an implementation of this tool and an evaluation of it on an IaC benchmark dataset.
Our evaluation shows that the tool scales well for modern IaC projects and runs quickly enough to be usable in practice.